# CartoAgent: a multimodal large language model-powered multi-agent cartographic framework for map style transfer and evaluation

Chenglong Wang[ab], Yuhao Kang[cd], Zhaoya Gong[ab]*, Pengjun Zhao[ab], Yu Feng[e], Wenjia Zhang[f], and Ge Li[g]

[a] *School of Urban Planning and Design, Peking University Shenzhen Graduate School, Shenzhen, China;*

[b] *Key Laboratory of Earth Surface System and Human-Earth Relations of Ministry of Natural Resources of China, Peking University Shenzhen Graduate School, Shenzhen, China;*

[c] *GISense Lab, Department of Geography and the Environment, University of Texas at Austin, Austin, USA;*

[d] *Department of Geography, University of South Carolina, Columbia, USA;*

[e] *School of Engineering and Design, Technical University of Munich, Munich, Germany;*

[f] *School of Architecture and Urban Planning, Tongji University, Shanghai, China;*

[g] *School of Electronic and Computer Engineering, Peking University Shenzhen Graduate School, Shenzhen, China*

* corresponding author: Zhaoya Gong, z.gong@pku.edu.cn

# CartoAgent: a multimodal large language model-powered multi-agent cartographic framework for map style transfer and evaluation

The rapid development of generative artificial intelligence (GenAI) presents new opportunities to advance the cartographic process. Previous studies have either overlooked the artistic aspects of maps or faced challenges in creating both accurate and informative maps. In this study, we propose CartoAgent, a novel multi-agent cartographic framework powered by multimodal large language models (MLLMs). This framework simulates three key stages in cartographic practice: preparation, map design, and evaluation. At each stage, different MLLMs act as agents with distinct roles to collaborate, discuss, and utilize tools for specific purposes. In particular, CartoAgent leverages MLLMs' visual aesthetic capability and world knowledge to generate maps that are both visually appealing and informative. By separating style from geographic data, it can focus on designing stylesheets without modifying the vector-based data, thereby ensuring geographic accuracy. We applied CartoAgent to a specific task centered on map restyling—namely, map style transfer and evaluation. The effectiveness of this framework was validated through extensive experiments and a human evaluation study. CartoAgent can be extended to support a variety of cartographic design decisions and inform future integrations of GenAI in cartography.

## 1. Introduction

Maps have significantly shaped human understanding of space and are recognized as one of the most important geographic ideas that have changed the world (Hanson, 1997). Cartography, the study and practice of making and using maps, has long been defined as a mixture of art, science, and technology (Krygier, 1995). For decades, cartographers have been exploring the use of cartographic language (such as symbols, annotations) to design maps that are both visually appealing and capable of effectively conveying accurate information with efficient map-making tools (Andrews, 1990; Li & Huang, 2002; Morrison, 1997; Nikmah et al., 2022). The development of information and communication technologies has significantly enhanced the richness of map-making tools and benefited the cartographic process. This evolution spans from traditional handcrafted maps, such as historical atlases, to digital mapping using geographic information system (GIS) software, and further to interactive web mapping platforms like OpenStreetMap[1]. Recently, the emergence of Generative AI (GenAI) has demonstrated significant potential across various applications (Mai et al., 2025; Messeri & Crockett, 2024; Milano et al., 2023; Thirunavukarasu et al., 2023) and opened up new opportunities to cartography (Ai, 2021; Kang et al., 2024; Zhang & Zhu, 2024). GenAI, exemplified by text-to-text large language models (LLMs) such as ChatGPT[2], Gemini[3], and LLaMA[4], as well as image generation models like Stable Diffusion[5], DALL·E[6], and Midjourney[7], can now interact naturally with users and generate artistic images per users' requests. Several scholars have explored the potential of GenAI for map creation (Christophe et al., 2022; Courtial et al., 2023; Kang et al., 2019; Schnürer et al., 2023; Tao & Xu, 2023; Wang et al., 2024b; Wu et al., 2024; Zhang et al., 2024a; Zhang et al., 2024b; Zhu et al., 2024) and discussed related ethical issues (Kang et al., 2023; Lin & Zhao, 2024; Zhao et al., 2021).

---

[1] https://www.openstreetmap.org/

[2] https://openai.com/chatgpt/

[3] https://gemini.google.com/app

[4] https://llama.meta.com/

[5] https://stability.ai/

[6] https://openai.com/index/dall-e-3/

Despite its promise, we identify two research gaps in the use of GenAI models for cartography. First, while LLMs can assist in the map-making process (Zhang et al., 2024a; Zhang et al., 2024b), the artistic side of cartography, which encodes human creativity and cartographers' experience (Caquard & Griffin, 2018; Tyner, 1987), has received less attention. Second, when image generation models are used to create maps, the raster-based outputs often lack both accuracy and informativeness. Accuracy refers to the extent to which a generated map correctly represents real-world geographic features and spatial relationships. Image generation models frequently produce fictitious regions that do not correspond to actual locations and distort spatial relationships between objects. Informativeness, on the other hand, measures how effectively a map conveys meaningful and interpretable geographic information. A map is considered informative if it clearly differentiates between object types, uses intuitive visual variables, and presents legible labels and icons that enhance understanding. However, image generation models often struggle with these aspects (Courtial et al., 2023; Kang et al., 2023; Lin & Zhao, 2024; Wu et al., 2022a; Wu et al., 2022b; Zhao et al., 2021). Therefore, these challenges pose the following research question: ***Is it possible to***

[7] https://www.midjourney.com/home

***develop a novel GenAI model that efficiently and automatically generates visually appealing maps while preserving both geographic accuracy and informativeness?***

The emergence of multimodal LLMs (MLLMs), particularly the groundbreaking GPT-4, has introduced a promising solution for various applications (Li et al., 2023; Lu et al., 2024; Wen et al., 2023). MLLMs are an advanced type of AI capable of understanding data from multiple modalities—such as images, text, audio, and video—and generating textual outputs (Wu et al., 2023). For example, researchers have explored the use of MLLMs to generate textual reports from medical images (Li et al., 2023). In map-making, MLLMs offer three key advantages. First, they possess strong text-processing capability that image generation models lack. During map design, geographic data and design styles are typically treated as separate components: geographic data record the precise locations of objects, while design styles dictate how these objects should be rendered. Cartographers specify visual variables in stylesheets using text rather than directly editing raster-based images. This practice ensures the ***accuracy*** of maps and could potentially be automated using multiple MLLMs. Second, MLLMs are pre-trained on vast amounts of text from the Internet and possess extensive world knowledge. When applied to map-making, these models can interpret the real-world meanings of map elements, thereby facilitating the creation of ***informative*** maps. Third, due to their visual aesthetic capability, MLLMs can analyze maps and provide feedback on layout and design. This helps create maps that are not only accurate and informative but also ***visually appealing***.

Moreover, several advanced autonomous systems have adopted a multi-agent approach powered by LLMs to tackle complex tasks (Akinboyewa et al., 2024; Li & Ning, 2023; Ning et al., 2024; Ning et al., 2025). In these systems, each agent acts as a goal-directed entity that can observe and interact with its environment (Wang et al., 2023). It possesses distinct capabilities and can collaborate with others to decompose tasks and solve problems (Guo et al., 2024). By leveraging such systems, we can improve the ***efficiency*** and ***automation*** of the cartographic process.

In this study, we introduce CartoAgent, a multi-agent cartographic framework powered by MLLMs. CartoAgent leverages the multimodal processing capability of MLLMs and the coordination capability of multi-agent systems to automate map creation. To validate the framework, we applied it to a task centered on the restyling aspect of map-making—map style transfer and evaluation. Specifically, we decomposed the map-making process into three key stages: preparation, map design, and evaluation. For each stage, we developed MLLM-powered agents specialized in different roles with clearly defined interactions, including an image appreciator, a stylesheet designer, an icon designer, a file implementer, and a map reviewer. Extensive experiments, including multi-scale and multi-source map style transfers, along with a human evaluation study, demonstrate the effectiveness of the framework. Taking advantage of the visual aesthetic capability and world knowledge of MLLMs, CartoAgent can create maps that are both visually appealing and informative. Additionally, by separating style from geographic data, CartoAgent focuses solely on designing stylesheets without modifying the vector-based data, which ensures the geographic accuracy of maps.

This article is structured as follows: In Section 2, we review the literature on map style transfer, map style evaluation, MLLMs, and autonomous agents. Section 3 presents the conceptual framework of our proposed CartoAgent. In Section 4, we instantiate our conceptual framework by applying it to map style transfer and evaluation, and provide key details of the technical implementation. Section 5 presents the experiments and their results. In Section 6, we analyze the results and discuss key takeaways. Finally, we conclude the article in Section 7.

## 2. Literature review

In this section, we first review existing research on map style transfer and its evaluation, and highlight their respective strengths and limitations. We then introduce MLLMs and autonomous agents, along with their current applications in GIS and cartography.

### *2.1 Map style transfer*

A map style is a collection of distinct yet aesthetically cohesive features in cartographic design (Kent & Vujakovic, 2017). Map style transfer refers to the process of reproducing a specific style—whether from an existing map or an artistic work—onto new map data (Kang et al., 2024). Previous studies have developed methods for map style transfer using both vector and raster data.

Vector data represents geographic entities as points, polylines, and polygons with precise locations and associated attributes. In vector map production, design styles are typically kept separate from geographic data. Cartographers have developed cartographic grammars that define style specifications and visual variables to guide map design decisions (Diamond, 2019; Griffin, 2021; Nestel, 2019; White, 2017). Researchers have used color pickers to extract color schemes from Monet's landscape paintings and applied these schemes to vector data for map style transfer (Friedmannová, 2009). Furthermore, Christophe and Hoarau (2012) highlighted the importance of considering color proportions and complementary principles in the process of map style transfer. Wu et al. (2022a) developed an optimization algorithm for transferring colors from arbitrary images to vector maps. However, previous studies have faced limitations in automating map style transfer. They have primarily focused on the color dimension while neglecting other elements such as size and shape, which are also important in map design. Moreover, they often overlook the semantic context of objects on the map, which makes it difficult for users to intuitively extract meaningful information. All these challenges highlight an opportunity to explore the integration of GenAI in map design.

A raster map typically consists of an array of pixels arranged in a rectangular grid, with each pixel storing numerical values that represent geographic information (Williams, 2019). The use of AI for map style transfer is rooted in raster-based maps. For example, researchers have developed advanced Generative Adversarial Networks (GANs) to transfer styles between map images within identical geographical contexts, as well as to apply artistic styles of famous paintings to maps (Bogucka & Meng, 2019; Kang et al., 2019; Christophe et al., 2022). Although these models show promise in manipulating the artistic aspects of cartography, they lack a deep understanding of spatial structures and real-world connotations of map objects. As a result, the raster-based outputs often contain blurred areas, fictitious regions, distorted relationships, and nonsensical symbols or annotations (Wu et al., 2022b; Kang et al., 2024), which makes it difficult for users to determine true geographic locations and extract meaningful information.

Therefore, our proposed CartoAgent aims to address these limitations. In line with common cartographic practices, we first separate style from geographic data to maintain accuracy. CartoAgent then incorporates two specialized agents that fully leverage the world knowledge and visual aesthetic capability of MLLMs. One agent, the stylesheet designer, is responsible for defining visual variables of objects in textual stylesheets, while the other, the map reviewer, evaluates and provides feedback on layout and design. This practice enables CartoAgent to automatically refine style specifications and ensures that maps are not only visually appealing but also accurate and informative.

### *2.2 Map style evaluation*

Map evaluation aims to enhance the quality and effectiveness of cartographic design through critical analysis (Bjørke, 1996; Robinson & Nelson, 2015). In particular, the evaluation of map styles should guide the map style transfer process to produce styled maps that are visually appealing, accurate, and informative (Wu et al., 2022b). Current metrics for evaluating map styles can be categorized into objective and subjective measures.

*Objective metrics.* Early explorations in this field were inspired by image quality assessment methods, where mathematical rules are defined to evaluate styled images. For example, some researchers introduced the structural similarity index from computer science to assess the effectiveness of styled map tile generation (Chen et al., 2020). More recently, researchers have begun integrating cartographic design constraints, such as smoothness and coalescence reduction, into the evaluation framework for styled maps (Courtial et al., 2022). Additionally, efforts have been made to directly develop AI models for map style evaluation. For instance, Kang et al. (2019) curated natural images and maps to train a binary classifier that determines whether distinct map characteristics are retained after map style transfer. These objective metrics are useful for conducting large-scale evaluations of styled maps. However, they have several limitations. On the one hand, due to the inherent complexity of map design, these metrics often provide only a superficial assessment. They usually focus on surface-level attributes like visual similarity, but fail to capture deeper and more critical aspects of cartographic quality, such as user perception and contextual relevance. On the other hand, scholars have pointed out that while machines can easily compute these metrics, cartographers may find them difficult to interpret and apply effectively in their decision-making processes (Kang et al., 2024).

*Subjective metrics.* Due to the limitations of objective metrics, researchers often turn to user evaluation studies. For instance, Chen et al. (2020) invited GIS/Cartography graduates to participate in a cognitive experiment where they evaluated and rated styled maps. The statistical measures derived from these evaluations (such as the average score) were used to assess the effectiveness of the corresponding approaches. Additionally, researchers have collected a large volume of assessment data from participants and trained models to simulate user preferences. For example, in evaluating the quality of sequential color schemes, Chen et al. (2021) first curated a dataset consisting of 781 five-color schemes, accompanied by assessments from 26 undergraduate students. An AdaBoost model was subsequently trained to align with the ratings provided by the undergraduates. Furthermore, Wang et al. (2024a) utilized Siamese neural networks to capture readers' perceptions of map information load, thereby guiding map design. While these metrics provide a more in-depth assessment, they are not without challenges. On the one hand, such studies require significant time and human resources. On the other hand, aggregating diverse perspectives and preferences from different experts may lead to an averaging effect, which could potentially overlooks individual differences in design preferences.

In this study, to address these challenges, we leverage visual aesthetic capability of MLLMs and their world knowledge to conduct map style evaluations. This approach enhances efficiency and enables in-depth assessments. Additionally, since MLLMs can utilize tools, they can perform objective measurements, such as comparing the visual similarity between images.

### *2.3 MLLMs and their applications in GIS/Cartography*

LLMs have demonstrated impressive capabilities across various tasks (Messeri &

Crockett, 2024; Milano et al., 2023; Thirunavukarasu et al., 2023). Researchers have been actively exploring different ways to leverage these models in the fields of GIS and cartography (Mai et al., 2022; Mai et al., 2024; Mai et al., 2025). For example, some studies have attempted to utilize LLMs and publicly available geospatial data to generate thematic maps or to construct mental maps from textual descriptions (Tao & Xu, 2023). Additionally, other researchers have developed methods to effectively instruct LLMs to perform spatial analysis or generate spatial database queries (Jiang & Yang, 2024; Wei et al., 2024; Zhang et al., 2024a; Zhang et al., 2024c; Zhang et al., 2024d).

More recently, MLLMs—extensions of LLMs that incorporate multimodal capabilities such as vision—have garnered significant attention (Wu et al., 2023). Researchers have enhanced the visual understanding of LLMs either by aligning these models with visual encoders through fine-tuning on image-text pairs (Zhu et al., 2023) or by leveraging visual language models to convert visual inputs into textual descriptions that these models can comprehend (Zeng et al., 2022). Prior studies have employed various methods, including multiple-choice questions and interactive chatbot arenas, to assess the perception, cognition, and analytical capabilities of MLLMs (Xu et al., 2024). Additionally, researchers have explored the potential of MLLMs in the field of remote sensing. For instance, Li et al. (2024) recognized the tremendous value that MLLMs could bring to visual tasks in remote sensing. Furthermore, the exploration of MLLMs' capabilities has extended to cartography, where researchers have tested their capabilities in reading and analyzing maps. Preliminary findings suggest that MLLMs can effectively extract information from maps (Xu & Tao, 2024; Zhang et al., 2025). However, these initial experiments primarily focus on map understanding, such as identifying objects on maps, while more in-depth investigations into how MLLMs can be integrated into map evaluation processes remain lacking. In this paper, we explore how MLLMs' visual aesthetic capability and world knowledge can be leveraged for map evaluation and ultimately facilitate the creation of visually appealing, accurate, and informative styled maps.

### *2.4 Autonomous agents and their applications in GIS/Cartography*

An autonomous agent is a system that interacts with its environment by perceiving its surroundings and taking actions over time to achieve specific goals (Wang et al., 2023). The development of such powerful agents has been a long-standing objective in both industry and academia. However, earlier efforts primarily focused on deploying models in isolated environments with limited knowledge, which greatly differed from human learning processes and hindered agents from making human-like decisions (Mnih et al., 2015). Recently, the integration of extensive world knowledge into LLMs and MLLMs has led to significant advancements in their capabilities and attracted considerable attention within the autonomous agent community. Researchers have begun exploring the use of LLMs and MLLMs as central controllers to build agents capable of human-like decision-making (Qin et al., 2023; Shen et al., 2023). Over time, a relatively unified framework—encompassing modules for profiling, memory, planning, and action—has emerged and been widely adopted (Wang et al., 2023).

Researchers have developed various autonomous agents, such as ChemCrow, which can assist in conducting chemical experiments (Bran et al., 2023), and ChatLaw, a legal AI consultant (Cui et al., 2023). This exploration has also extended into GIS and cartography research and led to a series of outstanding works. For example, Zhang et al. (2024a) introduced GeoGPT, a novel framework that leverages LLMs to understand user needs and autonomously invoke predefined GIS tools for geospatial data collection, processing, and analysis. Similarly, Li & Ning (2023) adopted LLMs as controllers and enabled them to automate spatial data collection, analysis, and visualization. In cartography, Zhang et al. (2024b) proposed the MapGPT framework, which integrates LLMs with multiple professional mapping tools. Users can interact with MapGPT through conversation to adjust map elements. These studies have significantly enhanced the automation of spatial analysis and visualization and lowered the barriers to using GIS software. However, one limitation remains: the artistic side of cartography, which encodes human creativity and cartographers' experience, is still largely underexplored. Agents often rely on human instructions or feedback to perform operations and refine their outputs. Previous studies have paid limited attention to autonomous agents designed to create visually appealing maps. In this paper, we integrate visual aesthetic capability and world knowledge of MLLMs with coordination capability of multi-agent systems to generate high-quality, styled maps that are visually appealing, accurate, and informative.

## 3. Conceptual framework

In this section, we first outline major steps in the map-making process. Next, we introduce our conceptual CartoAgent framework and discuss how it contributes to the intelligent map-making.

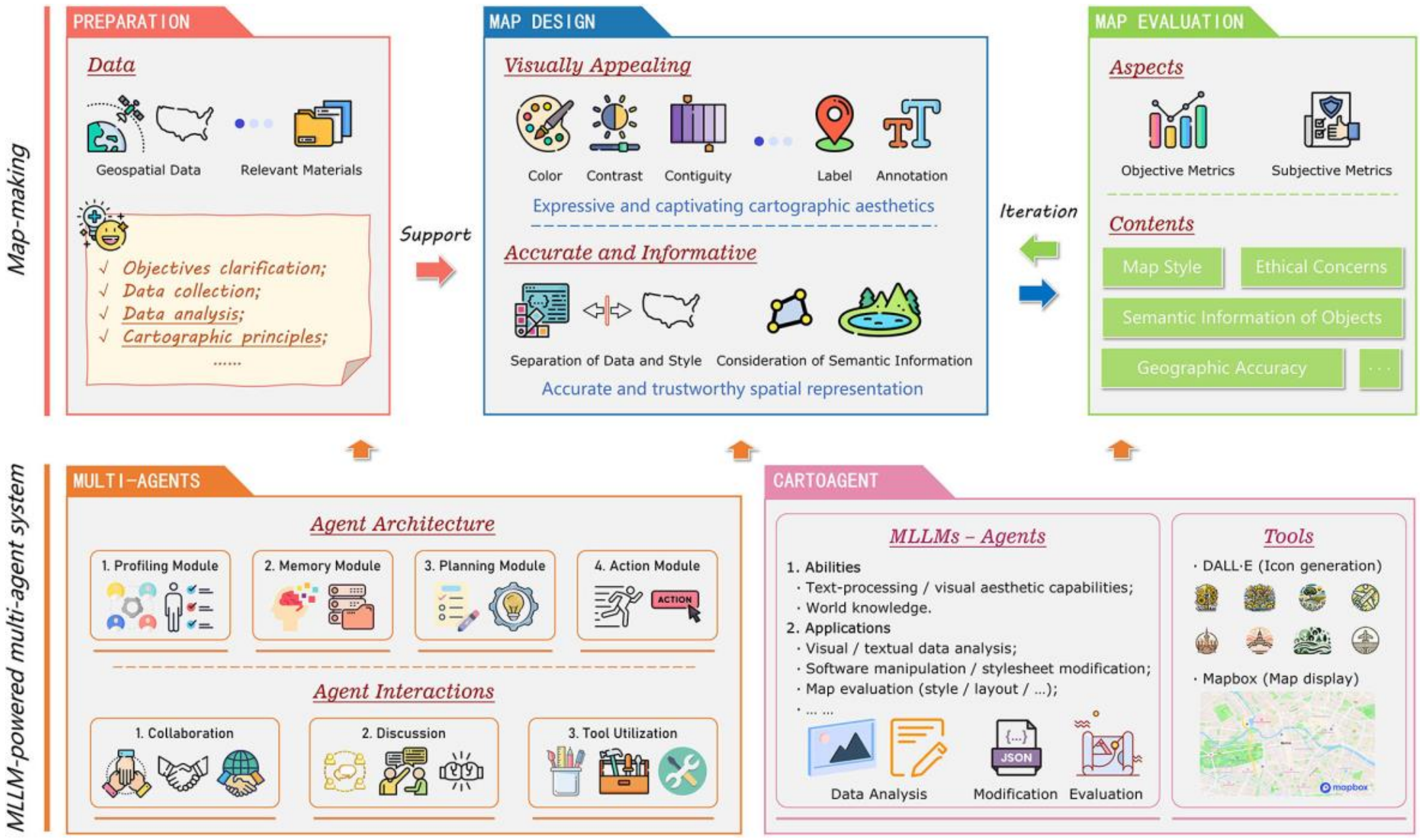

Figure 1. A conceptual cartographic framework powered by MLLMs.

### *3.1 Major steps in the map-making process*

The process of map-making involves several stages, including conceptual design, iterative prototyping, and comprehensive evaluation (Roth, 2019). In line with standard cartographic practices, we distill the process into three fundamental steps (Figure 1): preparation, map design, and map evaluation.

*Preparation.* In this initial step, cartographers define their objectives and carefully analyze geographic data along with other relevant materials. It is essential for them to consider fundamental cartographic principles and choose appropriate cartographic language to convey information effectively. Keeping map readers in mind is crucial, as it ensures that geographic information is accurately interpreted and users can form a reliable representation of the real world through maps.

*Map design.* Once the objectives are clarified and the datasets are prepared, the map design process begins. Cartographers often separate style from geographic data. They manipulate visual variables such as color, contrast, and contiguity within stylesheets and incorporate symbols and annotations to enhance visual appeal. Manipulating visual variables does not involve editing the geographic data itself, which can preserve spatial relationships between objects and ensure geographic accuracy. Additionally, since the function of a map is to communicate information to readers, a good map should be intuitive and easy to read. Therefore, cartographers often consider real-world connotations of objects when designing maps—for example, using cool tones to represent bodies of water.

*Map evaluation.* After map design, the next step is to assess the quality and effectiveness of maps and conduct map critiques. Map reviewers analyze whether the visual variables—such as size, color, and shape—are appropriately configured and whether the map as a whole is visually appealing. Additionally, they evaluate whether the cartographic design decisions make the maps accurate and informative.

It should be emphasized that the '*map design–map evaluation*' process may need to be repeated multiple times. Cartographers often adjust visual variables and refine the visual hierarchy iteratively to ensure the final design effectively meets the map's intended objectives. Following this conceptual framework, we introduce CartoAgent—an MLLM-powered multi-agent system designed to support each step with specialized agents.

### *3.2 CartoAgent: an MLLM-powered multi-agent cartographic system*

Given the strong capabilities of MLLMs in both text processing and visual aesthetics, as well as their extensive world knowledge, we can design multiple MLLMs with distinct roles to handle specific subtasks in the cartographic process. For instance, one MLLM could focus on analyzing relevant materials, while another could be designed to generate a textual stylesheet. During the map evaluation stage, an MLLM could assess the quality and effectiveness of maps, as well as critique their style and layout. By deploying a series of MLLMs, we can develop a multi-agent autonomous system for map-making, which we term CartoAgent. Figure 1 provides an overview of these agents, including their architectures and interactions. Notably, a specific CartoAgent is introduced in Section 4 for map style transfer and evaluation. However, the conceptual framework discussed here is adaptable and can be leveraged to support a wide range of cartographic design decisions.

*Agent architecture.* In line with common practices in agent-related studies, our proposed CartoAgent consists of four modules: a profiling module, a memory module, a planning module, and an action module. In CartoAgent, the profiling module is implemented by crafting prompts that clarify roles of agents (e.g., a stylesheet designer, a map reviewer) and outline their specific map-making skills and responsibilities (e.g., data analysis, map design, and evaluation). For instance, we can initialize a stylesheet designer to create textual stylesheets and a map reviewer to assess the quality and effectiveness of maps. The memory module preserves content gathered from various inputs. In CartoAgent, this module primarily determines whether to memorize map styles that have been entered into the system and whether to retain feedback on previously styled maps. For example, when we prompt the stylesheet designer to modify the style of maps, the previous stylesheets can be stored in memory and referenced.

The planning module can be used to decompose and dynamically adjust tasks, similar to how humans adapt to their environment. It employs techniques such as chain-of-thought (Wei et al., 2022), tree-of-thought (Yao et al., 2023), and graph-of-thought (Besta et al., 2024) for effective task decomposition and adaptation. In CartoAgent, this module efficiently breaks down the cartographic process into manageable subtasks and simplifies complex cartographic design decisions. For example, the agent can divide map evaluation into two key dimensions: objective and subjective assessments. During the objective evaluation process, the agent can analyze the color histogram of styled maps by invoking external tools. Finally, the action module translates agents' decisions into concrete actions. For example, by clearly defining the action sets corresponding to each decision (e.g., external APIs), we can ensure that CartoAgent effectively executes multiple tasks across the three stages of map-making.

*Agent interactions.* Each agent is designed to handle a specific subtask in the cartographic process. CartoAgent, as a multi-agent system, facilitates various interactions among these agents, including collaboration, discussion, and tool utilization. Collaboration involves multiple agents with unique capabilities and expertise working together to tackle complex tasks. In CartoAgent, one agent may specialize in designing stylesheets, while another may focus on analyzing relevant materials. These agents collaborate to ensure that the map's style aligns with relevant materials. Discussion refers to multiple rounds of communication between agents to refine outcomes. Agents review each other's outputs and provide feedback to improve map design. For instance, in CartoAgent, through multiple interactions between the map design agent and the map critique agent, the system can iteratively enhance the quality and effectiveness of the maps. Beyond interactions among agents within CartoAgent, some agents may also utilize external tools to assist in completing their tasks. For example, an agent might use GIS software like Mapbox to visualize map data (Figure 1). Properly specifying these agent interactions is crucial for the system to successfully execute complex tasks.

## 4. Using CartoAgent for map style transfer and evaluation

In this section, we implemented the proposed CartoAgent framework for map style transfer and evaluation to demonstrate its applicability in supporting the restyling aspect of map-making. Specifically, this task consists of two main steps. The first step is to apply the artistic styles of inspiration sources to the map data. The second step is to evaluate the quality and effectiveness of the styled map and provide feedback. To accomplish this overall task, we developed five distinct agents: an image appreciator, a stylesheet designer, an icon designer, a file implementer, and a map reviewer. Each agent has unique responsibilities and can handle specific subtasks. For the convenience of further discussion, we define $X$, $Y$, and $Z$ as the inputs, intermediate results, and outputs of the system, respectively. Additionally, we use $R$ and $T$ to represent the role of each agent and the external tools they might use. This task can be abstracted as follows: Given an inspiration source ($X_{inspiration}$) and map data ($X_{data}$), a series of agents with different roles design and generate an appropriate stylesheet ($Y_{style}$) for rendering the map data. The styled map ($Z_{map}$) should not only closely resemble the style of the input inspiration source but also be accurate and informative.

### *4.1 Role assignment and interaction specification*

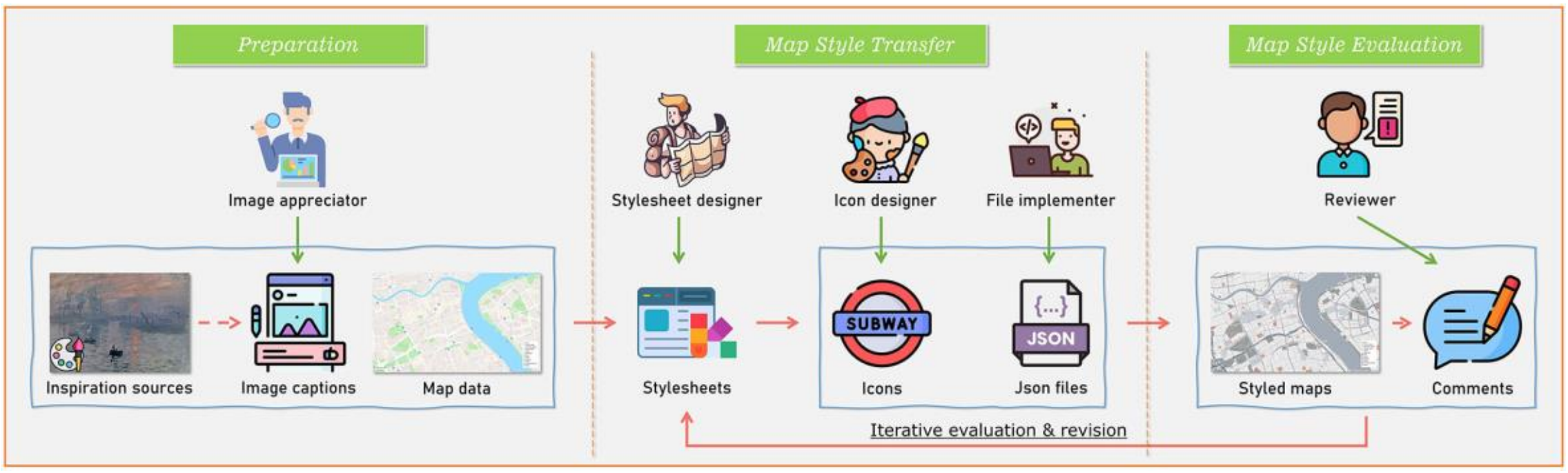

Figure 2. Role assignment and interaction specification.

Based on the conceptual framework, we divided the task of map style transfer and evaluation into three stages: preparation, map style transfer, and map style evaluation. Throughout these stages, we deployed five distinct roles: an image appreciator, a stylesheet designer, an icon designer, a file implementer, and a map reviewer. The overall workflow is illustrated in Figure 2, and we provide the pseudo-code for map style transfer and evaluation in Table 1.

Table 1. Pseudo-code for map style transfer and evaluation.

| **Algorithm: Map style transfer and evaluation** |
|---|
| $\boldsymbol{procedure}\ transfer - evaluation(X_{inspiration}, X_{data})$ |
| $Y_{captions} = R_{appreciator}(X_{inspiration})$ |
| $Y_{style} = R_{style_designer}(X_{inspiration}, X_{data}, Y_{captions})$ |
| $\boldsymbol{while}\ True$: |
| $\quad Y_{icon} = R_{icon_designer}(T_{DALL\cdot E}(Y_{style}))$ |
| $\quad Y_{file} = R_{implementer}(Y_{style})$ |
| $\quad Z_{map} = T_{platform}(Y_{icon}, Y_{file})$ |
| $\quad Y_{comment} = R_{reviewer}(X_{inspiration}, Z_{map})$ |
| $\quad \boldsymbol{if}\ Y_{comment} == 'Accept'$: |
| $\quad\quad \boldsymbol{return}\ Z_{map}$ |
| $\quad \boldsymbol{else}$: |
| $\quad\quad Y_{style} = R_{style_designer}(Y_{style}, Y_{comment})$ |

*Preparation.* A thorough understanding of the inspiration source is a prerequisite for performing map style transfer. To achieve this, we introduced an image appreciator ($R_{appreciator}$) responsible for comprehensively analyzing images and generating detailed image captions ($Y_{captions}$).

*Map style transfer.* At this stage, we first initialized an MLLM-powered stylesheet designer ($R_{style_designer}$) to create the stylesheet based on the inspiration source, map data, and the image captions generated by the image appreciator. Next, an MLLM-powered icon designer ($R_{icon_designer}$) was customized to design symbols ($Y_{icon}$), and a file implementer ($R_{implementer}$) converted the stylesheet into a concrete file format ($Y_{file}$) recognizable by specific GIS platforms ($T_{platform}$). Notably, different GIS platforms use various formats for style storage (for example, QGIS[8] uses Styled Layer Descriptors, while Mapbox[9] uses a Style Specification File). Therefore, a file implementer is essential to handle these different file formats.

*Map style evaluation.* After importing the symbols and the file into the GIS platform, the styled map is displayed. To further enhance its quality and effectiveness, we initialized an MLLM-powered map reviewer ($R_{reviewer}$). This agent can provide comments ($Y_{comment}$) on whether the styled map preserves the style of its inspiration source, assesses whether the cartographic design decisions make the map accurate and informative, and offers specific, actionable suggestions for the stylesheet designer to refine the outcome. This iterative process of style refinement and map evaluation may continue through multiple rounds until the reviewer is satisfied with the final result.

### *4.2 Prompt design*

To ensure that agents can successfully complete their assigned subtasks, we should craft well-designed prompts in a conversational setting. These prompts should initialize the agents, clarify their roles, and specify their individual responsibilities. When constructing these prompts, we followed specific principles outlined by Ekin (2023).

- *Purposefulness.* At the beginning of a conversation, we should first clarify the purpose, as a clear purpose leads to a more informative dialogue.

---

[8] https://qgis.org/

[9] https://www.mapbox.com/

- *Conciseness.* We should use clear and concise language that directly relates to the topic, rather than misleading words and unnecessary jargon, as these hinder understanding and undermine effective communication.
- *Explainability.* We should enable the model to explain its own actions, as this will promote transparency and help us understand the rationale behind its responses.
- *Normativity.* We should standardize the output format of the model. A standardized output structure promotes consistency, organization, and clarity in conversations.

Based on these principles and considering the specific subtasks assigned to different roles, we designed prompts for the five agents accordingly.

**Image Appreciator**

I want you to act as an image appreciator. I will provide you with an image, and your task is to describe it in as much detail as possible. This means the following topics should be covered:

- Content: What is in the image? What are the attributes and relationships of these elements? What objects cannot possibly appear in this image? What temporal and spatial information does the image convey?
- Color: What are the colors of the objects and the background? Please provide hexadecimal color values as accurately as possible (e.g., #000000). What effect does the combination of these colors have on the brightness and contrast of the image?
- Theme & Design: If the provided image is a painting, what is the theme of the painting? (e.g., Realistic, Surreal) What are the characteristics of the design? (e.g., Simple, Detailed, Complex)

If you understand the requirements of this task, you can reply 'OK' and provide your description after I provide the image.

Figure 3. The prompt designed to initialize the image appreciator.

*Image appreciator.* The agent was deployed to analyze various visual inputs, including photographs, paintings, AI-generated maps, and web maps. Based on the principles of image appreciation (Oppenlaender, 2022), we instructed the agent to focus specifically on content, color, theme, and design (as illustrated in Figure 3):

- *Content.* The agent should understand complex spatiotemporal context information, accurately identify objects in images, analyze their respective attributes, and comprehend the relationships between them.
- *Color.* The agent should analyze the overall contrast and visual hierarchy of the image and be able to provide hexadecimal color values for each object as accurately as possible.
- *Theme & design.* The agent should assess the theme of the image (e.g., realistic, surreal) as well as the characteristics of the design (e.g., simple, detailed, complex).

*Stylesheet designer.* The agent was developed to create the stylesheet based on the inspiration source, map data, and image captions. First, we need to clarify two key principles of stylesheet design in the prompt: the styled map should be both informative and visually appealing. In other words, this cartographic design decision should enable readers to easily extract information—for example, by using different colors to distinguish various types of data, selecting intuitive color palettes, and ensuring that labels and icons are clear. Additionally, the styled map should maintain a design as consistent as possible with the inspiration source. Next, we should list the visual variables that can be adjusted, such as size and color (Table 2). Finally, we need to define a set of requirements that should be followed during the stylesheet design process (as illustrated in the specific prompt in Figure 4):

- *Completeness.* The agent should configure the visual variables of all objects on the maps without any omissions.

**Stylesheet Designer**

I want you to act as a stylesheet designer. I will provide you with map data and a reference image along with its description (content, color, theme, and design). Your task is to create a suitable stylesheet to render the map data. When designing the stylesheet, you should consider two aspects:

- **Informative:** Before you dive into the design, ensure you thoroughly understand the map data. The map represents a specific environment, so the data and its semantic information should guide your design decisions. Specifically: 1) The styles should be informative. For example, use colors strategically to differentiate between different categories of data. 2) Make sure the colors you choose are intuitive. For instance, use shades of blue for water, greens for vegetation, and browns or grays for mountains and urban areas. 3) Labels and icons should be legible and clear, without distracting from the functionality of the map.
- **Visually appealing:** Ensure the styled map is visually appealing. The style of the rendered map should be as consistent as possible with the reference image. Also, when choosing a style, please consider the contrast and visual hierarchy of the map to ensure it is easy to interpret.

Below are the map data and variables that you can modify:
**[Map data]:**

- Icon elements: ____;
- Label elements: ____;
- Line elements: ____;
- Fill elements: ____;
- Background.

**[Variables]:**

- For each icon element, you can describe the expected style in as much detail as possible, e.g., its content, color, theme, and design. The icon designer will design this icon according to your expectations;
- For each label element, you can set the text color and the text halo color;
- For each line element, you can set the line opacity and the line color;
- For each fill element, you can set the fill opacity, the fill color, and the fill outline color;
- For the background, you can set the background color.

Meanwhile, please pay attention to the following requirements:

1. **Completeness:** Please set the styles for all elements;
2. **Explainability:** Please clearly explain the reason for your choices;
3. **Normativeness:** All colors should be provided in hexadecimal format (e.g., #000000), and opacity should be a decimal between 0 and 1 (where 0 is transparent and 1 is opaque). Please strictly follow the following JSON format to return the results and only return this JSON file: {"reasoning": ..., "stylesheet":{"symbol (icon)":{"...": {"explanation": ..., "expectation": ...}}, "symbol (label)":{"...": {"explanation": ..., "text-color": ..., "text-halo-color": ...}}, "line":{"...": {"explanation": ..., "line-opacity": ..., "line-color": ...}}, "fill":{"...": {"explanation": ..., "fill-opacity": ..., "fill-color": ..., "fill-outline-color": ...}}, "background": {"explanation": ..., "background-color": ...}}}

If you understand the requirements of this task, you can reply 'OK' and think step-by-step, providing your reasoning and stylesheet after I provide the reference image and its description.

Figure 4. The prompt designed to initialize the stylesheet designer.

- *Explainability.* The agent should explain its cartographic design decisions, including the manipulation of specific visual variables, to help cartographers understand the rationale behind its actions.

- *Normativeness.* The agent should follow a standardized output structure, such as providing color values in hexadecimal form and ensuring that transparency is represented as a decimal between 0 and 1.

Table 2. The visual variables that can be adjusted in our experiment.

| Elements | Visual variables |
|---|---|
| Icon | Content, color, theme & design |
| Label | Text-color and text-halo-color |
| Line | Line-opacity and line-color |
| Polygon | Fill-opacity, fill-color, and fill-outline-color |
| Background | Background-color |

The text-based stylesheet, designed by the stylesheet designer, was then handed over to the icon designer and the file implementer for the generation of raster-based symbols and the implementation of the specific style file, respectively.

**Icon Designer**

I want you to act as an icon designer. I will provide you with the referent and expectations for the icon design. Your task is to consider the characteristics of the referent and follow my expectations for the icon design. Please ensure that each image contains only one icon, and the background is pure white.
[Referent]: ____;
[Expectations]: ____.

Figure 5. The prompt designed to initialize the icon designer.

*Icon designer*. This agent was deployed to generate symbols for point elements based on the expectations set by the stylesheet designer and the characteristics of the referent. We designed the prompt shown in Figure 5 to initialize the agent and leverage advanced image generation capability of GenAI models to create these symbols.

**File Implementer**

I want you to act as a file implementer. I will provide you with an original style specification file and a new stylesheet. Your task is to update the style specification file based on the new stylesheet, while strictly adhering to the format of the original file. Please ensure that the final result includes the entire modified style specification file, with the unmodified parts left intact. If you understand these instructions, please reply with 'OK' and provide the modified style specification file once I send you the original file and the new stylesheet.

Figure 6. The prompt designed to initialize the file implementer.

*File implementer*. Different GIS platforms use various formats for style storage. For example, QGIS uses Styled Layer Descriptors, while Mapbox uses a Style Specification File. Therefore, we initialized a file implementer to convert textual stylesheets into a concrete file format recognizable by specific GIS platforms. The prompt is shown in Figure 6.

**Map Reviewer**

I want you to act as a map reviewer. I will provide you with two images: one is a reference image, and the other is a map designed to match the style of the reference image. Your task is to evaluate the quality of the styled map based on two criteria:

- **Informative:** Does the map design consider the semantic information of the objects? Are different types of data distinguished by different colors? Are the chosen colors intuitive? Are the labels and icons legible and clear, without distracting from the map's functionality?
- **Visually appealing:** Is the style of the map consistent with the reference image, considering aspects such as color, theme, and design?

Once you have completed the evaluation, you can make one of the following choices:

- **Accept:** If you think the map is good as it is, you don't need to do anything else;
- **Revision:** If you think improvements are needed, provide a modified stylesheet and detailed explanations. Note that you don't need to modify all map data, only what's necessary.

Our map data includes: ____.

- For each icon element, you can specify the expected style as precisely as possible (e.g., content, color, theme, and design). The icon designer will design this icon based on your specifications;
- For each label element, you can modify the text color and the text halo color;
- For each line element, you can adjust the line opacity and line color;
- For each fill element, you can change the fill opacity, fill color, and fill outline color;
- For the background, you can change the background color.

Please adhere to the following requirements:

1. All colors should be provided in hexadecimal format (e.g., #000000);
2. Opacity should be a decimal between 0 and 1 (where 0 is fully transparent and 1 is fully opaque).

Please use the following JSON format to return your final results:

- If you choose 'Accept', return this JSON: {"Action": "Accept"}
- If you choose 'Revision', return this JSON: {"Action": "Revision", "Modified stylesheet": "..."}

If you understand the task, please reply 'OK' and provide your answer after I provide the reference image and the styled map.

Figure 7. The prompt designed to initialize the map reviewer.

*Map Reviewer.* To enhance the quality and effectiveness of the styled map, we introduced a map reviewer designed to mimic how cartographers critique maps. This agent can evaluate whether the styled map is informative and aligns well with the style of the inspiration source. If the agent is satisfied with the final result, it directly accepts the map; otherwise, it requests a revision and provides suggestions for further improvement (Figure 7).

## 5. Experiments and results

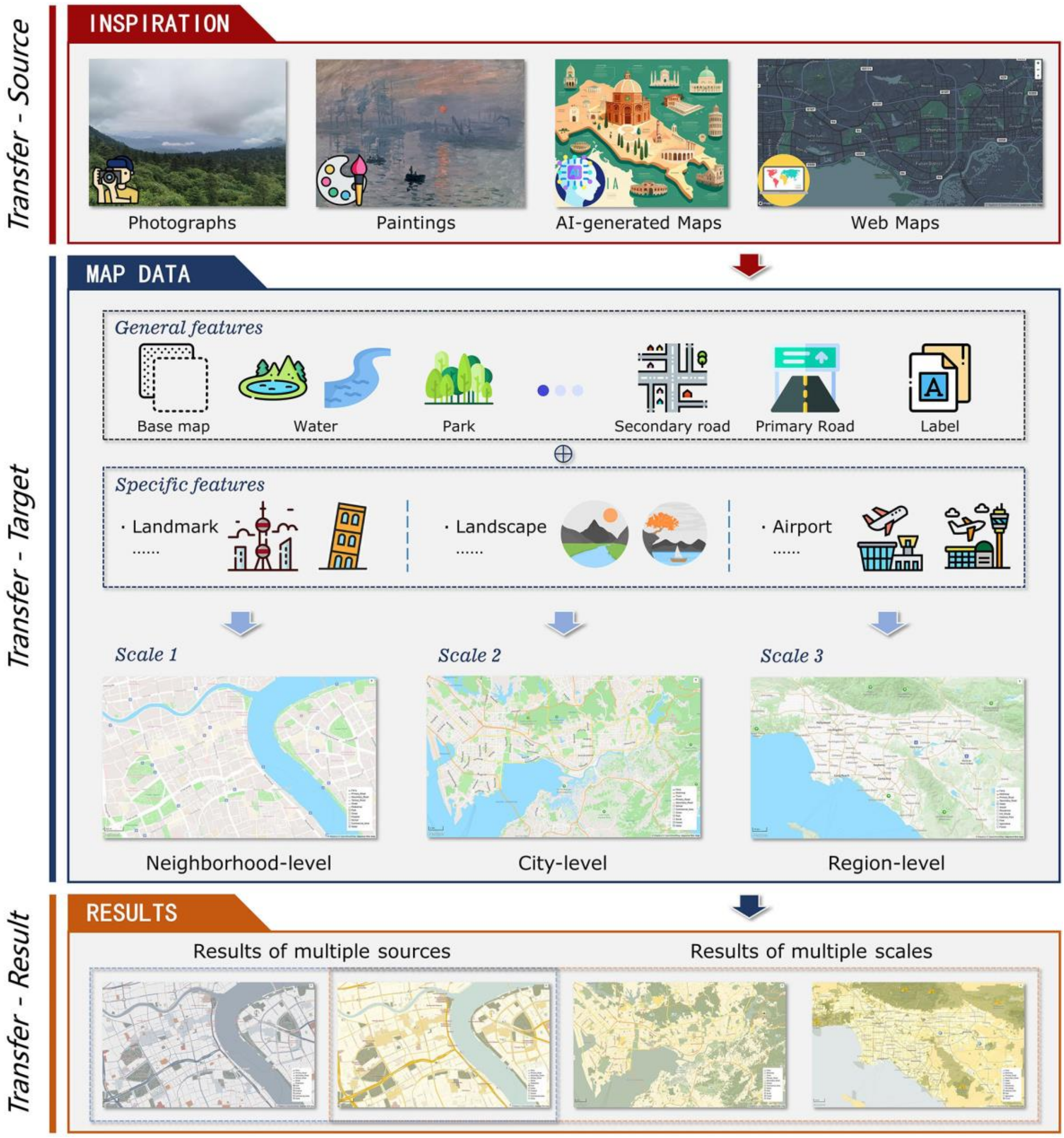

Figure 8. Design of experiments.

In this section, we demonstrate the applicability of our CartoAgent in map style transfer and evaluation using GPT-4o, DALL·E, and the Mapbox platform. We first introduce the design and implementation details of our experiments. Next, we present the results of multi-scale and multi-source map style transfers, with an in-depth analysis of a specific case: transferring the style of *Sunflowers* to map data. Finally, we recruit 17 experts and students and assess the capabilities of MLLMs in map style evaluation through a human evaluation study.

### *5.1 Design of experiments and implementation details*

Table 3. Multi-scale map data.

| Category | Features | | |
|---|---|---|---|
| | **Neighborhood-level** | **City-level** | **Region-level** |
| Point (icon) | 1. Oriental pearl tower<br>2. Metro station | 1. Natural landscape<br>2. Cultural landscape | 1. Airport<br>2. National park |
| Point (label) | 1. Natural landscape<br>2. Cultural landscape<br>3. Primary road<br>4. Secondary road<br>5. Tertiary road<br>6. Street road | 1. Natural landscape<br>2. Cultural landscape<br>3. Primary road<br>4. Water | 1. Airport<br>2. National park<br>3. Major settlement<br>4. Minor settlement |
| Line | 1. Ferry<br>2. Primary road<br>3. Secondary road<br>4. Tertiary road<br>5. Street<br>6. Pedestrian | 1. Ferry<br>2. Motorway<br>3. Trunk<br>4. Primary road<br>5. Secondary road | 1. Ferry<br>2. Motorway<br>3. Primary road<br>4. Secondary road |
| Polygon | 1. Park<br>2. Grass<br>3. Hospital<br>4. School<br>5. Commercial area<br>6. Water | 1. School<br>2. Commercial area<br>3. Grass<br>4. Park<br>5. Scrub<br>6. Forest<br>7. Water | 1. Water<br>2. Airport<br>3. Residential<br>4. Hill shade<br>5. National park<br>6. Park<br>7. Agriculture<br>8. Forest |

*Design of experiments.* Figure 8 provides an overview of the design of experiments for our map style transfer and evaluation case study. We selected various types of inspiration sources for the experiments, including photographs, paintings, AI-generated maps, and web maps. Photographs and paintings are frequently used as inspiration sources in map style transfer studies (Wu et al., 2022b). In addition, researchers have noted that while AI-generated maps[10] may not be directly usable, they possess visual appeal and can serve as vital inspiration sources in the map-making process (Kang et al., 2023). Furthermore, our study incorporated web maps, which are often presented online as tiled raster-based images. Consequently, it becomes challenging and time-consuming to extract the underlying stylesheets and replicate their styles.

[10] https://aballatore.space/2023/01/16/ive-made-some-maps-using-ai/

As shown in Table 3, we conducted experiments using multi-scale map data. Maps at different spatial scales emphasize diverse features in their representations due to map generalization. For instance, regional-level maps highlight long-distance transportation infrastructure, such as airports. These experiments serve to thoroughly validate the feasibility of the proposed CartoAgent framework. Specifically, the data comes from Mapbox and includes all the information provided by the Mapbox standard map at a specific zoom level within a defined region.

*Implementation details.* We utilized the GPT-4o and DALL·E models to develop the multi-agent system and integrated the Mapbox platform to display styled maps. Specifically, DALL·E, which generates images from textual descriptions, was used to initialize the icon designer, while the other roles were initialized using GPT-4o. The resulting icons and style file were then imported into the Mapbox Style Editor to create customized maps. However, since the Style Editor does not support the addition of map elements such as a north arrow or scale bar, we obtained the style URL from the editor and built a simple web page using Mapbox GL JS to incorporate these components. Although this required some manual effort—a common issue for most GenAI-assisted design tasks—it accounted for only a small portion of the overall process. The system is flexible, not limited to specific models or platforms, and can be easily extended. All experiments were conducted through the official interactive interface. It is important to note that before each map style evaluation, we cleared the map reviewer's memory.

This approach prevents the reviewer from lowering its standards due to prior interactions or shared memory with other agents.

### *5.2 Results of multi-scale and multi-source map style transfers*

Figures 9–11 present styled maps based on various inspiration sources and multi-scale map data, with additional examples provided in Appendix B. In these figures, the left column displays the inspiration sources, the middle column shows the styled maps without evaluation, and the right column presents the styled maps after evaluation.

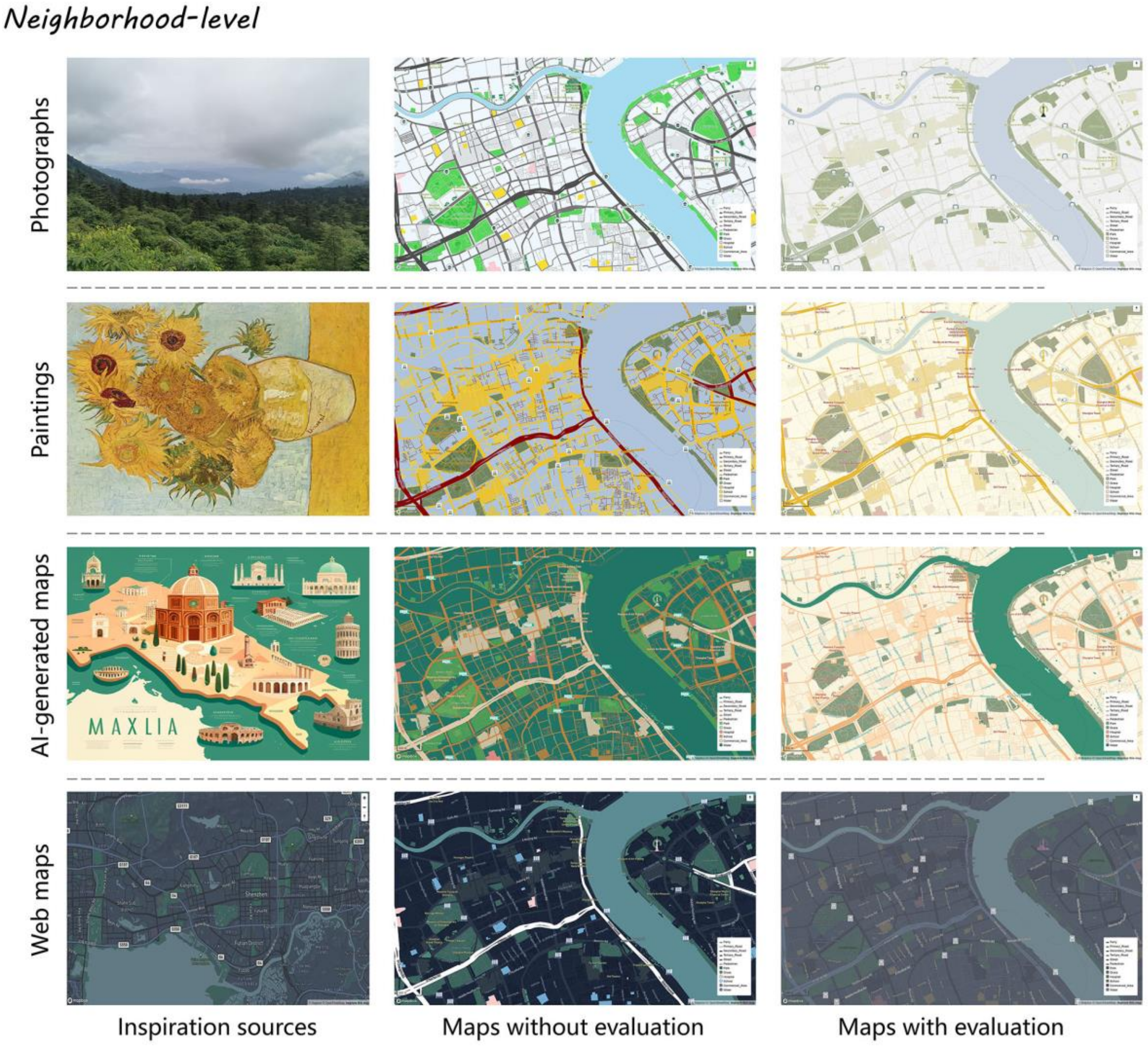

Figure 9. Examples of neighborhood-level map style transfers.

We make the following observations: On the one hand, although the colors in the pre-evaluation maps sometimes resemble those in the inspiration sources, there are still noticeable differences in the specific color values. For instance, in Figure 9, although the photograph prominently features a dark green tone, GPT-4o assigned a bright green color to parks. Similarly, while the web map has a dark gray background, GPT-4o instead used a bright black tone. These differences in color choices reduce the stylistic similarity between the styled maps and their inspiration sources.

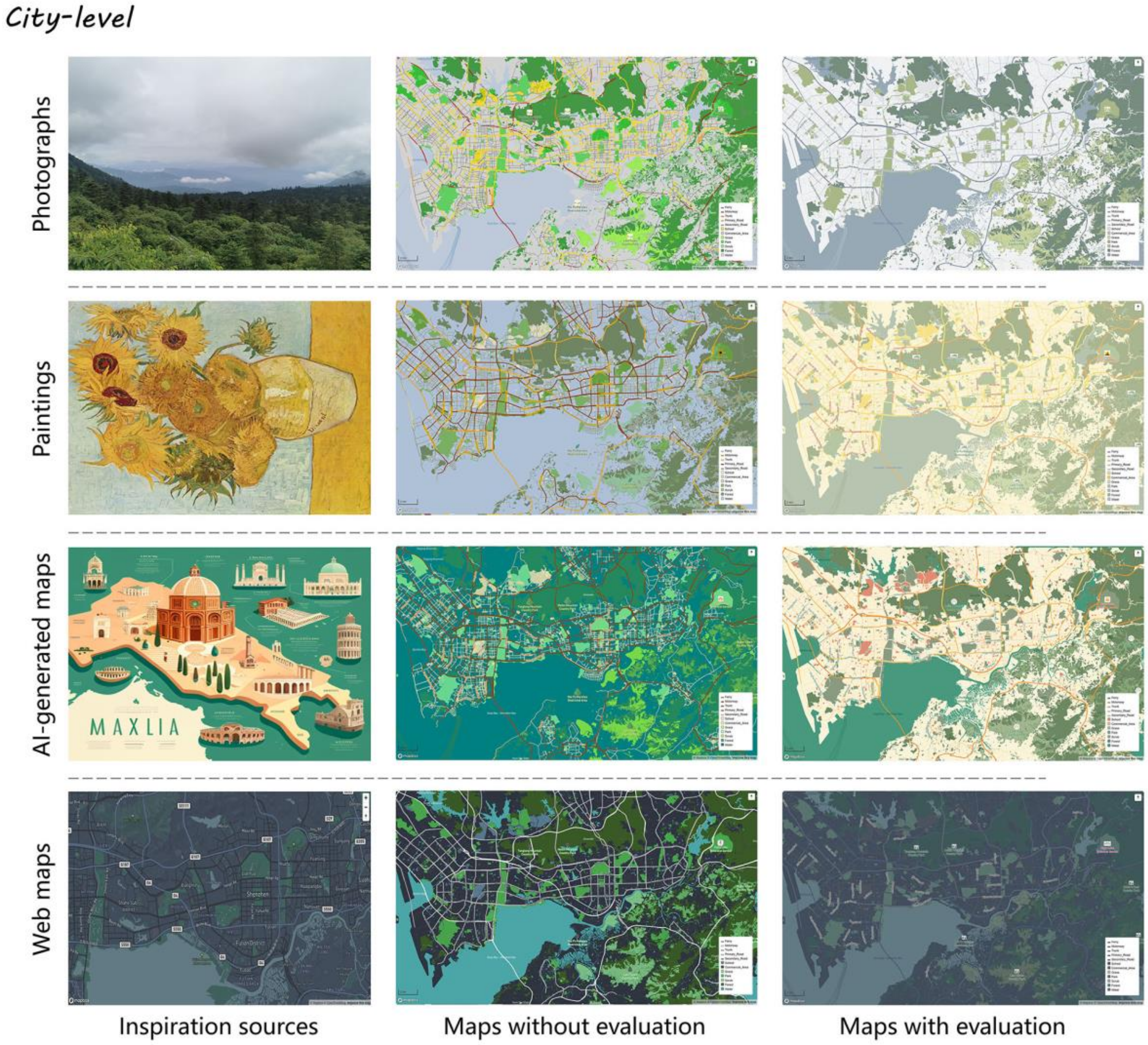

Figure 10. Examples of city-level map style transfers.

To quantify these differences, we converted all RGB images to the HSV color space, which better reflects human color perception. We then computed the cosine similarity between the color distributions of the maps—either unevaluated or

evaluated—and those of the inspiration sources. As shown in Figure 12a, regardless of how the HSV space is discretized into bins, map evaluation consistently improves the similarity to the inspiration sources. When we used only 10 bins, the improvement is limited—likely because the coarse binning fails to capture subtle color differences, especially when the pre-evaluation maps already resemble the sources to some extent. However, when the hue, saturation, and value channels were each divided into 20 bins, the evaluation significantly enhances color similarity. As illustrated in Figures 12b, 12c, and 12d, the similarity score can increase dramatically—from 0.2217 to as high as 0.7900.

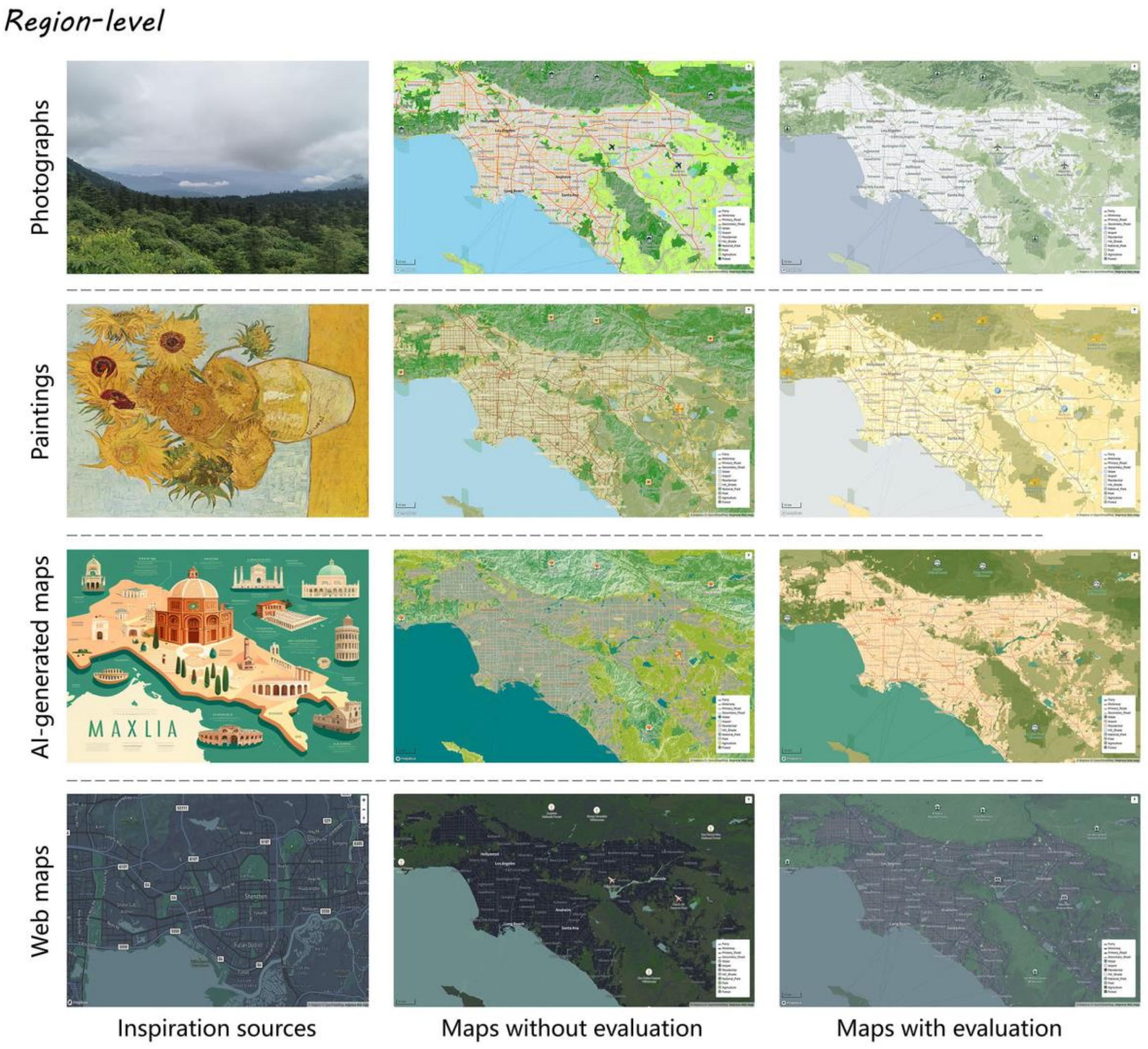

Figure 11. Examples of region-level map style transfers.

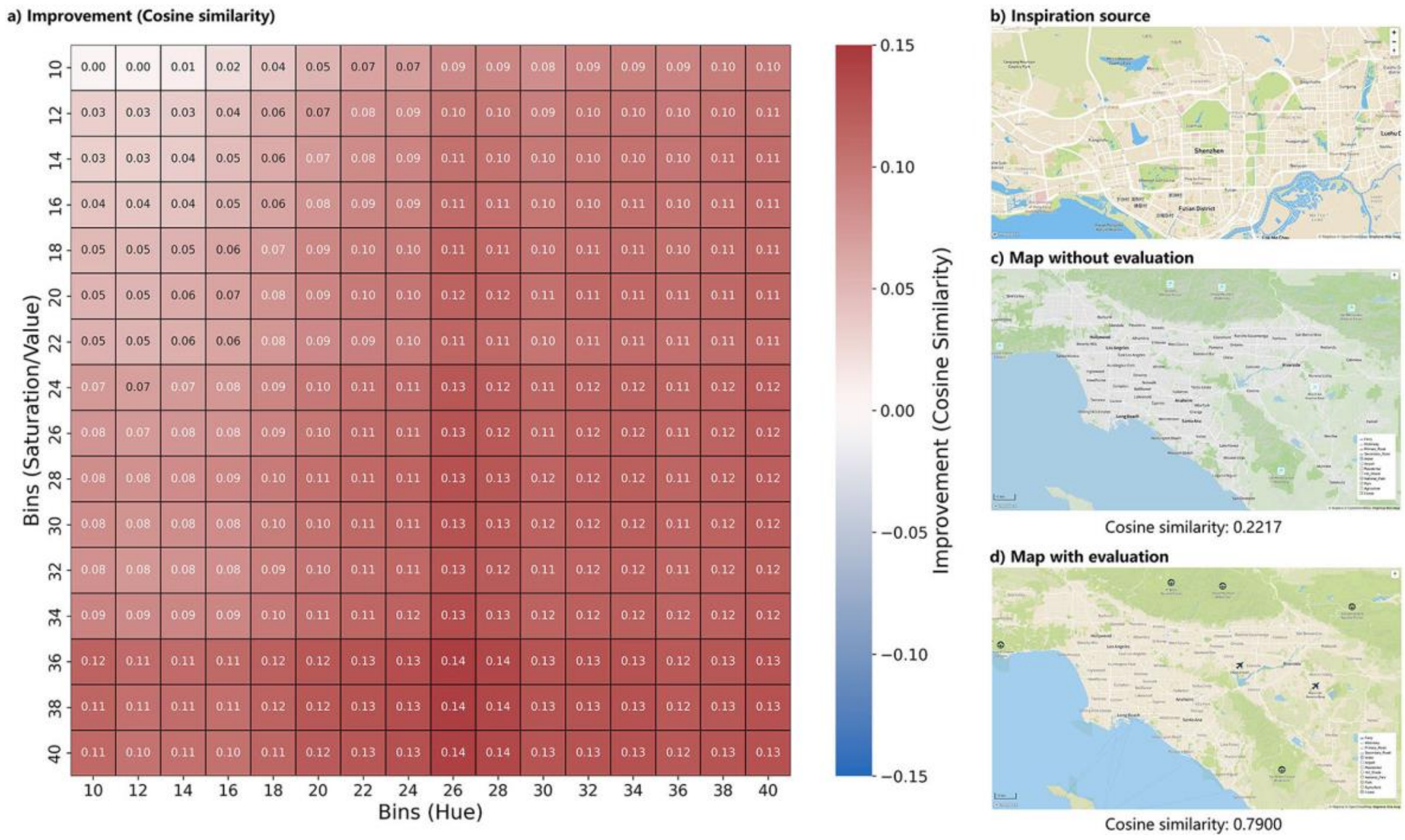

Figure 12. The improvement in cosine similarity with map evaluation.

On the other hand, maps without evaluation are often uninformative. For example, in Figure 9, two pre-evaluation maps set the background and water to the same color, which makes these elements indistinguishable. This design choice compromises figure-ground separation and results in a suboptimal visual experience. In contrast, the styled maps with evaluation not only effectively capture the distinctive styles of the inspiration sources but also consider contrast, visual hierarchy, and the real-world connotations of elements, which ensures a high level of communicative quality.

Moreover, by separating style from geographic data, CartoAgent focuses solely on restyling the maps without editing the vector-based data. This approach ensures the accuracy of the geographic representation. Statistically, these maps underwent an average of 12.63 design and evaluation iterations between the map reviewer and the stylesheet designer. Notably, this iterative refinement process mirrors the steps followed by cartographers to achieve optimal design solutions.

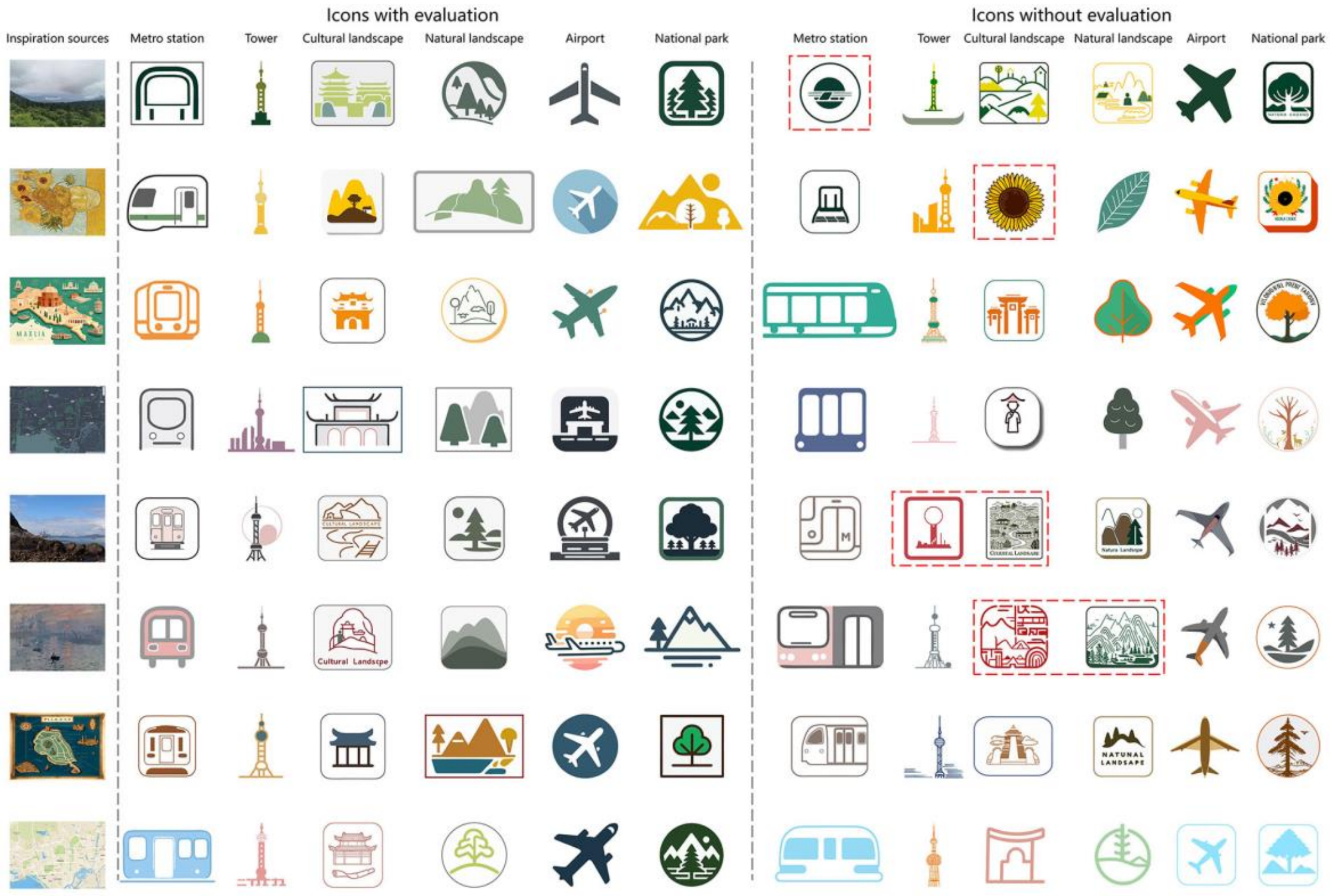

Figure 13. Icons generated by the icon designer within our multi-agent system.

Figure 13 illustrates the icons generated by our multi-agent system, which effectively meet users' personalized needs. We make the following observations: First, icons created without evaluation are often suboptimal—some are overly complex, which makes them unclear when displayed on a map, while others are unintuitive and difficult for readers to understand. Second, after evaluation, the icons become clearer and easier to interpret. They not only reflect color similarities to their inspiration sources but also stand out more prominently on the map. Third, icons generated by DALL·E still have certain flaws. For instance, the model occasionally generated hallucinated text and produced incorrect labels like 'Cultural Landscpe' instead of the correct 'Cultural Landscape.'

### *5.3 A detailed case study: transferring the style of Sunflowers to map data*

In this section, taking *Sunflowers* as an inspiration source, we provide an in-depth analysis of the system's workflow and performance. In *Sunflowers*, Van Gogh captures the vibrant beauty of nature by skillfully blending warm tones. The interplay of colors and textures presents a significant challenge for style transfer.

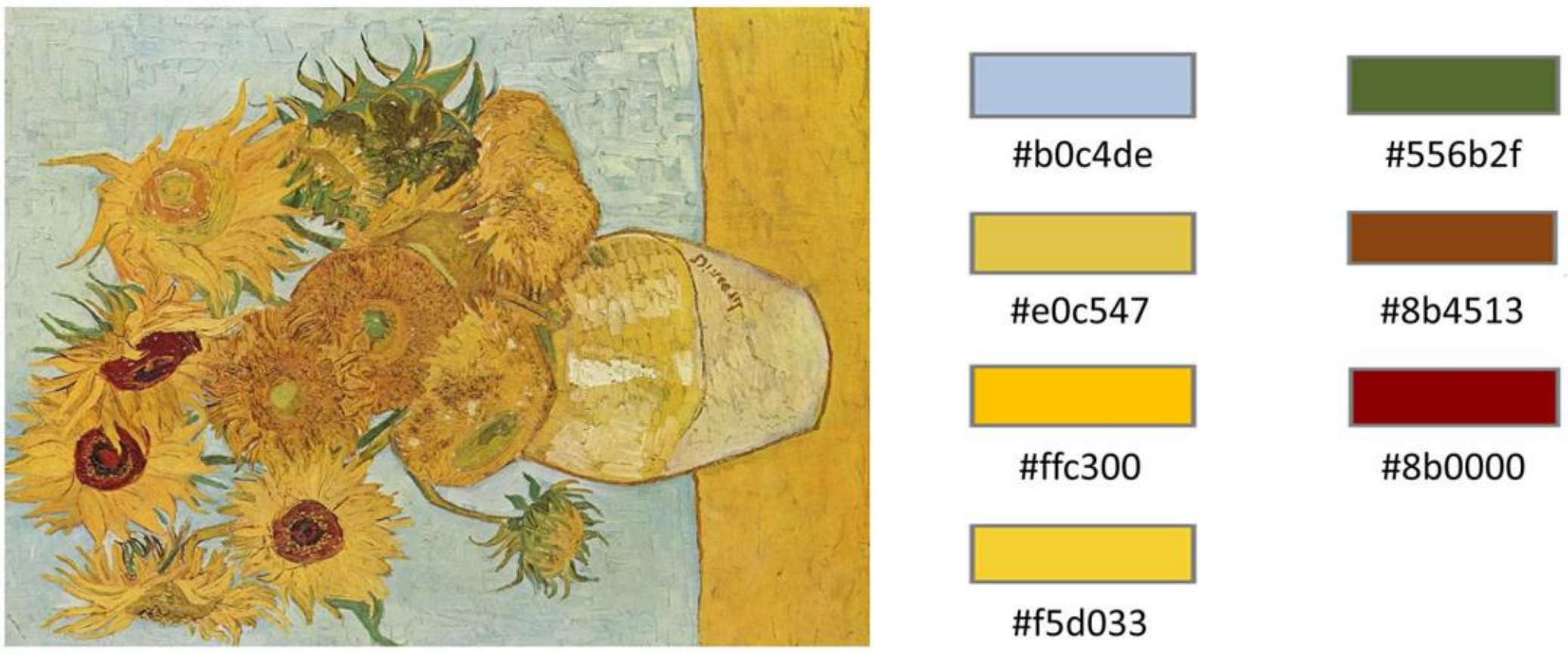

**Content**

The image is a painting that depicts a vase filled with sunflowers. The sunflowers are prominently placed and appear to be in various stages of bloom and wilt. The vase, which is quite simple in design, serves as the central object holding these vibrant flowers. The background is relatively plain with a soft texture, allowing the flowers to be the focal point of the painting.

- **Attributes:**
  - **Sunflowers:** They have large, round, yellow petals with some showing dark brown or reddish centers. Some sunflowers appear to be wilting, indicating different states of life.
  - **Vase:** A simple ceramic vase with a light, somewhat muted yellow color, adds subtlety, so as not to distract from the flowers.
  - **Background:** It is a cool, pale blue (#b0c4de), creating a calm atmosphere.
  - **Surface:** The table or podium is a warm yellow (#e0c547), adding to the sunflower's vibrancy.
- **Relationships:** The sunflowers are the primary subjects, arranged in a way that displays their natural beauty and imperfection. The vase acts as a grounding element, containing the flowers but not overpowering them. The background provides a contrasting hue that makes the yellow tones of the sunflowers stand out more vividly.
- **Impossibilities:** Objects like modern electronics, animals, or humans cannot appear in this image as it focuses entirely on the still life of flowers in a vase.
- **Temporal and Spatial Information:** The painting suggests an indoor setting, likely a quiet and serene space where the focus is on the natural beauty of flowers. The style and items in the painting suggest a time period before modern technology, possibly late 19th century, as indicated by the painter's technique and choice of subject.

**Color**

The predominant colors in the painting are yellows (#ffc300, #e0c547, #f5d033), greens (#556b2f), and a pale blue background (#b0c4de).

- **Sunflowers:** Different shades of yellow dominate the petals, including golden yellow (#ffc300) and darker mustard (#e0c547). The centers of some flowers are brownish (#8b4513) or red (#8b0000).
- **Vase:** The vase features light yellows (#f5d033) and some reflections creating a blend of creamy and tan hues.
- **Background:** A soft blue (#b0c4de) that brings a calming contrast to the sunflowers warm tones.
- **Table:** The surface is a warm and slightly muted yellow (#e0c547).

The combination of the warm, vibrant yellows of the sunflowers and vase against the cool, serene blue background adds to the brightness and contrast of the image. The use of complementary colors (yellow and blue) enhances the visual separation of the subject from the background, making the sunflowers appear more vivid and lively.

**Theme & Design**

- **Theme:** The painting focuses on capturing the natural beauty of sunflowers.
- **Design Characteristics:**
  - **Detailed:** The brushwork is expressive yet captures the intricate details of the petals and centers of the flowers.
  - **Complex:** The arrangement of flowers varies in size, angle, and state of bloom, adding complexity and depth to the composition.
  - **Textural Elements:** The background and vase are rendered with somewhat simplified texture, helping to accentuate the detailed sunflowers.

The painting's design emphasizes natural beauty, contrasting elements, and an appreciation for the ephemeral nature of life, as suggested by the wilting flowers among the blooming ones. This still life painting is an iconic representation of Vincent van Goghs style, known for its bold colors and dynamic brushwork.

Figure 14. Image captions generated by the image appreciator.

Figure 14 presents the image captions generated by our image appreciator. This agent successfully identified key objects in the scene, such as sunflowers, a vase, the background, and the surface. More impressively, it recognized the spatiotemporal context of the painting. It described the environment as quiet and serene, with a time period likely predating modern technology—possibly the late 19th century. Additionally, the agent accurately provided hexadecimal color values and offered a thoughtful interpretation of the artwork's theme and design.

```json
{
    "reasoning": "The reference image is a painting of sunflowers in a vase by Vincent van Gogh. To create a map stylesheet that reflects the color, theme, and design of this painting, I will use vibrant yellows for significant elements, muted blue for the background, and natural greens for vegetation areas. This approach aligns with the color palette and the focus on nature present in the painting. Attention will be given to ensure contrast and clear visual hierarchy, making the map both informative and visually appealing.",
    "stylesheet": {
        "line": {
            "Pedestrian": {
                "explanation": "Pedestrian paths should be subtly shown but clear, hence using lighter shades.",
                "line-opacity": 0.8,
                "line-color": "#e0c547"
            },
            "Street": {
                "explanation": "Streets should have a medium visibility with warmer tones to reflect general urban areas.",
                "line-opacity": 0.9,
                "line-color": "#8b4513"
            },
            "Tertiary_Road": {
                "explanation": "Tertiary roads should be easily distinguishable without overwhelming the map.",
                "line-opacity": 0.85,
                "line-color": "#f5d033"
            },
            "Secondary_Road": {
                "explanation": "Secondary roads are of moderate traffic importance and should be clearly delineated.",
                "line-opacity": 0.9,
                "line-color": "#ffc300"
            },
            "Primary_Road": {
                "explanation": "Primary roads must stand out prominently for easier navigation.",
                "line-opacity": 1.0,
                "line-color": "#8b0000"
            },
            "Ferry": {
                "explanation": "Ferry lines on water should use cooler tones but still be visible.",
                "line-opacity": 0.7,
                "line-color": "#556b2f"
            }
        },
        ... ...
    }
}
```

Figure 15. The stylesheet crafted by the stylesheet designer.

As shown in Figure 15, our stylesheet designer effectively generated high-quality stylesheets based on the inspiration source, map data, and image captions. The design not only considers the artistic aspects of cartography but also ensures that the styled maps remain informative. For example, when selecting a color for ferries, the agent chose a cool yet visible tone that aligns with real-world connotations and the inspiration source. Likewise, for roads, brighter tones were applied to primary and secondary roads to emphasize their importance, while more subdued colors were used for streets and tertiary roads to establish a clear visual hierarchy.

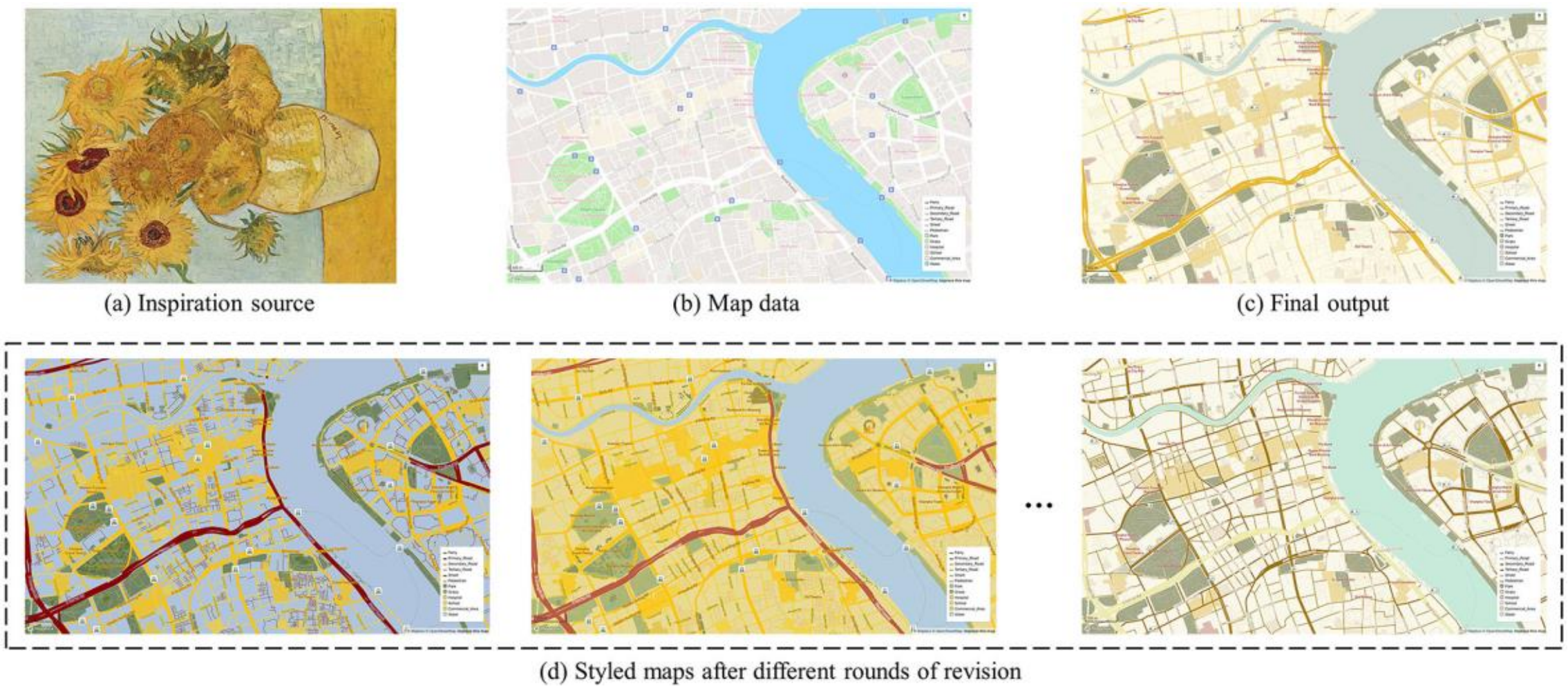

Figure 16. The styled maps after different rounds of revision.

After the icon designer and file implementer generated the symbols and created the style file based on the stylesheet, the styled map was displayed on the Mapbox platform. The map reviewer then evaluated whether the map is informative and whether its artistic style aligns with that of the inspiration source and provided detailed and actionable feedback for adjustments. This process of style refinement and map evaluation went through several iterations. As shown in Figure 16, after multiple rounds of revision, the map reviewer was ultimately satisfied with the result. Initially, the styled map exhibits overly high color saturation and poor figure-ground separation, which makes it inconsistent with the style of the inspiration source and difficult for readers to interpret. However, through continuous adjustments, the final version closely resembles the inspiration source and becomes much easier to understand. Moreover, because we separate style from geographic data and agents only manipulate the visual variables in the stylesheets without editing the data itself, there is no loss in accuracy.

Table 4. The detailed, actionable feedback for adjustment provided by the map reviewer.

| Map data | Suggestions | Explanations |
| --- | --- | --- |
| Natural landscape label | Text-color: #6B8E23<br>Text-halo-color: #FAF3D3 | Text color should reflect earthy tones, with a light halo for readability over complex backgrounds. |
| Primary road label | Text-color: #B74E3E<br>Text-halo-color: #FAF3D3 | Reduce the intensity of the red to a muted terracotta, for better harmony with the reference image. |
| Primary road | Line-color: #B74E3E<br>Line-opacity: 0.9 | Soften the road color to avoid overwhelming the map, using a warm brown-red. |
| Street | Line-color: #E3B55A<br>Line-opacity: 0.8 | Lighten the street color to blend with the yellow background without losing definition. |
| Park | Fill-color: #91A76F<br>Fill-opacity: 0.7<br>Fill-outline-color: #6B8E23 | Use a textured, muted green to reflect the organic shapes and earthy tones of the reference image. |
| Water | Fill-color: #AFCDE7<br>Fill-opacity: 0.9<br>Fill-outline-color: #87B0D9 | Tone down the blue to a softer, more pastel shade, echoing the sky background in the painting. |
| | … … | |
| Background | Background-color: #FAF3D3 | Shift the background to a warm yellow to reflect the canvas-like feel of the reference painting. |

Table 4 presents the actionable feedback provided by the map reviewer. To enhance figure-ground separation, the reviewer suggested adjusting the background color to a warm yellow and softening the blue of the water to a more pastel shade. For road labels, specific color adjustments were recommended: the color of primary roads should be changed to a warm brown-red to reduce their visual dominance, while the color of streets should be lightened to better harmonize with the yellow background without compromising clarity. These thoughtful recommendations showcase the reviewer's visual aesthetic capability. In the next section, we will further evaluate the reviewer's capability through a human evaluation study.

### *5.4 Human evaluation study*

Prior studies have emphasized the importance of involving humans in evaluating the results generated by GenAI (Chen et al., 2025; Voas et al., 2023; Zhuang et al., 2024). To further assess the visual aesthetic capability of MLLMs and ensure that the styled maps align with human preferences, we conducted a human evaluation study with 17 experts and students (10 male and 7 female) majoring in GIS and cartography (Figure 17). Specifically, we organized the results of 24 experiments in the form of questions. In each question, participants were shown a reference image (the inspiration source) along with the final styled map and three randomly selected intermediate styled maps. These maps were presented in a random order without labels indicating the number of revisions. Participants were then asked to select the best map based on two key aspects: the effectiveness of the design in conveying semantic information (e.g., using different colors to represent various data types, intuitive color choices, and clear visibility of labels and icons) and the alignment of the artistic style with the reference image in terms of color, theme, and overall design.

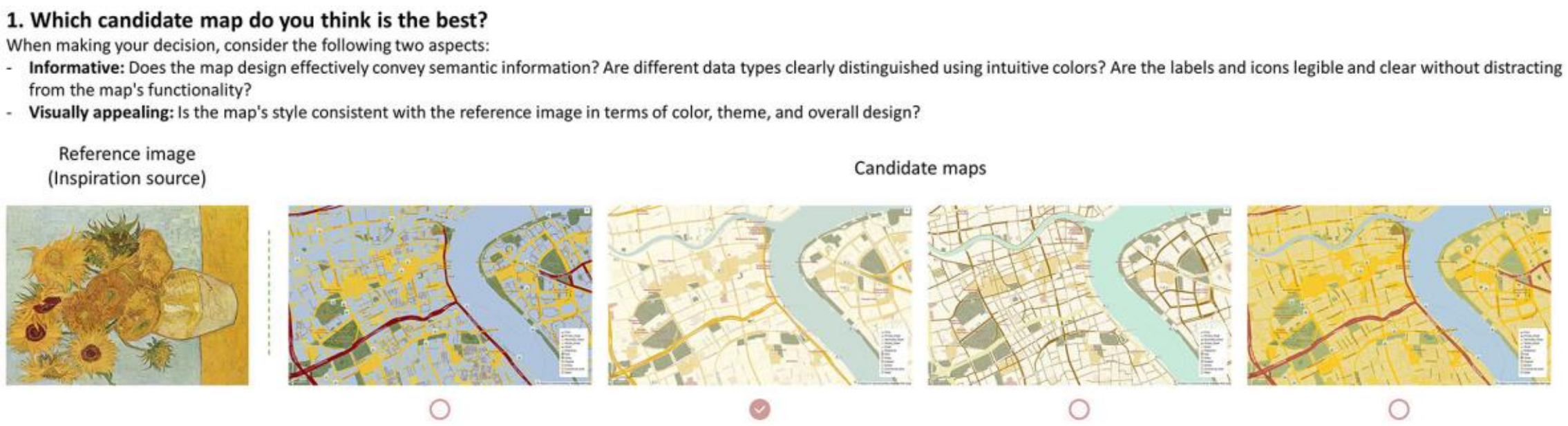

Figure 17. A question from the experiment, which illustrates how we conducted the human evaluation study.

By determining whether the maps selected by participants match the final styled maps accepted by the agent, we can assess the extent to which the MLLMs' evaluation criteria align with human judgment. Overall, the results are promising: if a map selected by participants matches the final styled map accepted by the agent, it is considered correct; otherwise, it is classified as incorrect. The accuracy reaches 83.82% ± 8.68%.

This suggests that MLLMs have the potential to leverage their visual aesthetic capability and world knowledge to assist humans in evaluating map styles.

## 6. Discussions

In this section, we list several takeaways from the study and discuss the potential of MLLMs in advancing cartography.

### *6.1 What knowledge is required for map style transfer?*

When using CartoAgent for map style transfer and evaluation, users only need to upload an inspiration source and map data to generate a styled map, without requiring any prior knowledge. Since map style transfer involves more than just artistic decisions—maps must be visually appealing, accurate, and informative—we have made every effort to fully leverage the knowledge embedded in MLLMs and integrate additional expertise into the agent workflow. This approach enhances both the quality and effectiveness of the styled maps.

Specifically, we fully utilize three types of knowledge. First, we integrate cartographic knowledge into the agent workflow. For example, cartographers typically separate style from geographic data and manipulate visual variables to achieve effective map design. Following this common cartographic practice, we incorporate a specialized agent, the stylesheet designer, responsible for defining the visual variables of objects in stylesheets through textual descriptions. This approach mitigates a prevalent issue in AI-generated maps, which often contain blurred areas, fictitious regions, distorted relationships, and nonsensical symbols or annotations—all of which compromise geographic accuracy (Wu et al., 2022b; Kang et al., 2023; Kang et al., 2024). Second, we leverage the world knowledge embedded in MLLMs. Since MLLMs are pre-trained on vast amounts of text, they inherently possess knowledge about various objects and their attributes. For instance, they understand that bodies of water, such as oceans and lakes, typically appear blue due to the absorption and scattering of light. By explicitly instructing the model to focus on the semantic information of maps, we can harness this knowledge to enhance the quality and effectiveness of map design. Third, we explore the aesthetic knowledge of MLLMs. Although aesthetics are subjective, and defining what makes a map beautiful is challenging (Jiang & Sui, 2014; Jiang & de Rijke, 2021; Kent, 2017), assessing the stylistic similarity between a given inspiration source and a styled map is relatively straightforward. Since MLLMs have been fine-tuned on text-image pairs, they can analyze an inspiration source, recognize its content, color schemes, themes, and design, and evaluate a map's contrast and visual hierarchy. By instructing the model to value visual appeal, we can activate this aesthetic awareness during map style transfer. These three types of knowledge contribute to the generation of styled maps that are visually appealing, accurate, and informative.

Despite these advancements, there is still room for improvement. For example, integrating more specialized cartographic knowledge—such as the information theory of cartography (Li et al., 2021)—or insights from neuroscience (Wu et al., 2025) could enhance map design and visual quality. Moreover, map styles are influenced not only by size, color, and shape but also by the topological organization and orientation of geometric elements, human cognition and emotions (Jiang, 2020; Jiang & de Rijke, 2023; Ortag, 2009). Future research should enable agents to adjust a broader range of visual variables and conduct more extensive cognitive experiments to deepen our understanding of how visual design influences map perception.

### *6.2 How to evaluate the results of map style transfer?*

Maps should prioritize and balance visual appeal, accuracy, and informativeness—fundamental principles that guide our evaluation of map styles. Unlike generating maps directly using image generation models in prior studies, our approach separates style from geographic data to ensure geographic accuracy. As a result, we have resolved common issues observed in prior studies, such as blurred areas, fictitious regions, distorted relationships, and nonsensical symbols or annotations (Wu et al., 2022b; Kang et al., 2023; Kang et al., 2024). To create styled maps that are both visually appealing and informative, we must consider these two aspects during the evaluation process. On the one hand, the styled maps should closely resemble their inspiration sources; on the other hand, they must effectively convey spatial information without causing visual confusion. Since both aspects are explicitly outlined in the prompt, the agent can adhere to them strictly.

Furthermore, to assess the ability of MLLMs in evaluating map styles, conducting a human evaluation study is essential (Chen et al., 2025; Voas et al., 2023; Zhuang et al., 2024). In this study, we recruited 17 experts and students to participate in a cognitive experiment. The results indicate that the evaluation criteria of MLLMs largely align with human judgment and suggest that MLLMs have the potential to leverage their visual aesthetic capability and world knowledge to assist humans in evaluating map styles.

However, there is still room for improvement. Due to the inherent trade-off between communicative quality and aesthetic appeal, map style transfer can be viewed as a max-max optimization problem (Wu et al., 2022a). Some styled maps may prioritize communicative clarity over strict visual resemblance to their inspiration sources, while others may closely mimic the styles at the expense of slight reductions in communicative quality. Future advancements could enhance user experience by generating multiple styled map options and allowing users to choose the one that best aligns with their preferences and needs. This approach would enable greater personalization in map-making and cater to diverse aesthetic and functional requirements.

### *6.3 How to build a robust multi-agent cartographic system?*

A well-designed multi-agent cartographic system should be reliable and resilient while maintaining optimal performance. To enhance its robustness, we have carefully adhered to several key design principles throughout its development.

- *Explainability.* We require MLLMs to provide necessary explanations during the cartographic process. These explanations can help cartographers understand the rationale behind the models' actions and reduce the "black box" nature of deep learning models.
- *Agents Interactions.* Just like humans, agents can make mistakes, such as misidentifying the theme of an inspiration source. To address this, we allow agents to actively engage in discussions. For example, map reviewers evaluate the quality and effectiveness of the styled maps and provide detailed, actionable suggestions to the stylesheet designers. The final styled maps are produced after multiple rounds of revision, which significantly reduces the impact of any single mistake within the system.
- *Review mechanism.* Each time we conduct a map style evaluation, we clear the map reviewer's memory. In other words, we initiate a new session, assign GPT-4o the role of map reviewer through a predefined prompt, and upload the inspiration source and styled map. This ensures that the agent reassesses the quality and effectiveness of the styled map from scratch without being influenced by previous comments or feedback. This approach prevents the reviewer from lowering its standards due to prior interactions or shared memory with other agents. Given that MLLMs may perceive the same map differently due to variations in their pre-training datasets, future work could incorporate multiple MLLMs to provide a more comprehensive evaluation. Additionally, combining human evaluation studies and human-in-the-loop processes could help ensure the outputs better align with human perceptions.

### *6.4 Ethical issues related to map style transfer*

This application addresses the public's demand for customized maps, as AI-generated maps are more personalized than those created using existing resources. However, ethical concerns, particularly regarding copyright issues, cannot be overlooked (Zhao et al., 2021; Kang et al., 2024). Several potential stakeholders may claim copyright in this process. First, the system builder, who configures the model and provides computational resources, may assert a claim to copyright ownership. Second, the user, who contributes to open-source map data and selects inspiration sources, could argue for copyright based on their creative input in shaping the final styled map. Third, the original artists behind these inspiration sources may also contend for copyright over their artistic creations. This complex interplay of contributions raises significant legal and ethical considerations in cartography. Although current technology may not fully resolve these copyright challenges, there is an urgent need for intellectual property frameworks and policies specifically designed for AI-generated maps.

## 7. Conclusions and future work

In this study, we explored how advanced GenAI can be leveraged to create visually appealing, accurate, and informative maps. Specifically, we proposed CartoAgent, a multi-agent cartographic framework powered by MLLMs. We applied this framework to a task centered on the restyling aspect of map-making—map style transfer and evaluation. To support the three stages of the task—preparation, map style transfer, and evaluation—five agents were designed with distinct roles: an image appreciator, a stylesheet designer, an icon designer, a file implementer, and a map reviewer. These agents collaborated, discussed, and utilized tools to generate and evaluate maps. By harnessing the visual aesthetic capability of MLLMs, along with their extensive world knowledge, CartoAgent can produce maps that are both visually appealing and highly informative. More importantly, this approach ensures geographic accuracy as it focuses solely on designing stylesheets without modifying the vector-based data. To evaluate the feasibility of our approach, we conducted extensive experiments on multi-scale and multi-source map style transfers. Additionally, we recruited 17 experts and students and conducted a human evaluation study to confirm that the results align with human perceptions. Designed for users without cartographic expertise, CartoAgent enables personalized map creation and makes high-quality cartographic design accessible to a broader audience.

The contributions of this study are multifaceted. First, we proposed CartoAgent, which integrates the visual aesthetic capability and world knowledge of MLLMs with the coordination capability of multi-agent systems to generate high-quality and styled maps. Second, we demonstrated how GenAI models can benefit the future of cartography in terms of both its artistic dimension and scientific informativeness. Third, our study contributes to the existing literature on map style transfer and evaluation.

Unlike earlier approaches that exhibited issues such as blurred areas, fictitious regions, distorted relationships, and nonsensical symbols or annotations (Wu et al., 2022b; Kang et al., 2023; Kang et al., 2024), our framework ensures precise geographic representation by separating style from geographic data. Our approach customizes map styles by manipulating visual variables directly, without editing the vector-based data.

Although CartoAgent has presently been applied to map style transfer and evaluation, it can be extended to support a wide range of cartographic design decisions. For instance, agents could be designed to automatically acquire data, select appropriate map projections, and generate thematic maps, rather than just general reference maps. We believe the proposed CartoAgent framework offers significant opportunities across diverse domains, such as disaster response, urban planning, land use, and climate studies.

## Data and codes availability statement

The map data used in this research was sourced from Mapbox (https://studio.mapbox.com/). To facilitate replication of our results as closely as possible—despite GPT-4o not being a deterministic model—we have provided step-by-step instructions in the appendix (attached as a supplementary file), along with the code available in our repository (https://github.com/Bonj0ur/MapStyleTransfer).

## Acknowledgments

The authors would like to express sincere thanks to the Editor-in-Chief, Prof. May Yuan, the Guest Editor, Prof. Gengchen Mai, as well as all anonymous reviewers for their insightful comments and suggestions.

## Funding

This work was supported by the Guangdong Basic and Applied Basic Research Fund Committee, Provincial General Project [2024A1515011924]; the National Natural Science Foundation of China [42471186]; the Shenzhen Science and Technology Program [KQTD20221101093604016, JCYJ20220818100810024]. Yuhao Kang is partially supported by the McCausland Innovation Fund at the University of South Carolina, and the Cartography and Geographic Information Society Rising Award.

## Disclosure statement

No potential conflict of interest was reported by the authors.